\documentclass[12pt]{article}
\usepackage{amssymb}
\usepackage{amsfonts}
\usepackage{amsmath}
\usepackage{amsmath,amsthm}
\usepackage{subfigure}
\usepackage{graphicx,epsfig, color }

\oddsidemargin 10mm \numberwithin{equation}{section}
\def\be{\begin{equation}}
\def\ee{\end{equation}}
\begin{document}
\begin{center}
{{\bf { Thermodynamics phase transition of Anti de Sitter
Schwarzschild scalar-tensor-vector-Black Holes}} \vskip 1 cm {
Hossein Ghaffarnejad \footnote{E-mail address:
hghafarnejad@semnan.ac.ir} and Elham Ghasemi Kordkheilee
        \footnote{E-mail address: e\_ ghasemi@semnan.ac.ir}} }\\
   \vskip 0.1 cm
   {\textit{Faculty of Physics, Semnan University, P.C. 35131-19111, Semnan, Iran} } \\
    \end{center}
\begin{abstract}
Instead of scalar tensor gravity models which is applicable for
description of cosmic inflation with unknown dark sector of
matter/energy, at presentense there are presented different
alternative scalar tensor vector gravities where meaningful
dynamical vector fields can support cosmic inflation well without
to use dark matter/energy concept.  One of these gravity models
was presented by Moffat which
 its modified Schwarzschild black hole solution is used to
study thermodynamic phase transition in presence of the AdS space
pressure in this article. To do so we obtained an equation of
state which asymptotically reaches to equation of state of ideal
gas for large black holes but for small scale black holes we
obtained a critical point at phase space where the black hole can
be exhibit with a phase transition at processes of isotherm and
isobaric. By looking at diagrams of the Gibbs free energy and the
heat capacity at constant pressure which are plotted versus the
temperature and the specific volume one can see an inflection
point which means that the phase transition is type of second
order.  In fact there is small to large phase transition for the
black hole which is equivalent to the Van der Waals liquid-gas
phase transition in ordinary thermodynamic systems. The phase
transition is happened below the critical point in phase space
when the gravitational charge of the black hole is equal to its
mass.
\end{abstract}
\section{Introduction}
General relativity (GR) is an elaborate theory of gravity which
successfully has the most correspondence to the experiments
\cite{Will, Berti, Abbott, Hees}, but nonetheless there are some
unresolved issues that can not been explained by GR. Therefore,
various kinds of theories have been developed as generalization of
GR in order to solve such problems
\cite{Heisenberg,Clifton,Weinberg}. By calculating velocity of
galaxies in the clusters Zwicky concluded that the gravitational
mass is more than the luminous matter
 \cite{Trimble}. Consequently, dark matter was defined, but it has not been detected yet and GR has failed to explain cosmological
 data without postulated non-baryonic dark matter. In fact we require
 modified gravity theories where the gravity tensor is coupled non-minimally by dynamical vector fields and so dark sector of matter/energy
 problem resolved by understandable dynamical vector fields.
  These models are called as scalar-tensor-vector gravity (STVG) models \cite{Moffat,Moffat1,Ghaffar0, Ghaffar300}. These theories
  explain the phenomena without requirement to introduce dark matter \cite{Brownstein} and successfully explain
  the rotation curves of galaxies
   \cite{Moffat1}, the dynamics of galactic clusters \cite{Moffat2}, the growth of structure \cite{Moffat3}
  and the cosmic microwave background (CMB) acoustic power spectrum \cite{Moffat4}.
  Also, Cai and Miao have analyzed the quasinormal modes of the generalized ABG black holes \cite{Ayon1, Ayon2, Ayon3, Ayon4} in STVG
  theory \cite{CaiMi}. One of us is studied some applications of the STVG
  model given by \cite{Ghaffar0,Ghaffar300} in the references
  \cite{Ghaffar2010,Ghaffar2015,GhaffarDeh2019, Ghaffar2019}. Discovery of evaporation of black holes in presence of
  interacting quantum matter fields by Hawking in 1974 \cite{Haw, Haw1} is inferred one that there is a closed relationship between
  the unknown quantum gravity theory and the black hole thermodynamics. After his novel paper many physicists encouraged to expand
  studies about the black hole thermodynamics while the quantum gravity
  theory is still maintained as unknown. For instance Davis began to study the phase
transition of black holes \cite{Davies} and it became noticed
since the Hawking discovered the black holes radiation
\cite{Hawking}. Bardeen investigated about the laws of black holes
thermodynamics  \cite{Bardeenn}.  Bekenstein introduced that the
black hole entropy should be a quarter of the surface area of the
horizon $S=\frac{A}{4}$ \cite{Bekenstein}. Hawking and Page
obtained a first order phase transition for the AdS Schwarzschild
black hole \cite{Hawkingpage} which is known as Hawking-Page phase
transition. The next motivation had been made by Chambline et al
in \cite{Chamblin1,Chamblin2} where they discovered a small-large
phase transition in AdS-RN black hole which is the same as Van der
Waals liquid-gas phase transition. These studies are encourage
others to study thermodynamics of different kinds of AdS black
holes in several gravitational theories: Kerr-AdS black hole
\cite{Altamirano} is studied in GR theory, AdS-Schwarzschild,
AdS-RN and Kerr-AdS black holes thermodynamics are studied in
modified f(R) gravity \cite{Chen}, Born-Infeld gravity
\cite{Zou1}, Gauss-Bonnet gravity \cite{Cai2,Wei,Zou2}, dilaton
gravity \cite{Dehghani,Stetsko1,Stetsko2}, Lovelock gravity
\cite{Dolan1,Frassino,Xu} or higher dimensions
\cite{Gunasekaran,Belhaj,Dolan2} respectively.
  In this regard, the effect of the positive cosmological parameter was not forgotten and the evaporation of the quantum black holes
   in the presence of a cosmological constant prevented the final destruction of
   the quantum black holes. In practice, the cosmic parameter was a restraining force (see for instance
    \cite{Ghaffar2,Ghaffar3,Ghaffar4}).
    In this sense modified laws of black holes thermodynamics
   were generated where fluid hydrodynamics behavior of the black holes are described by $VdP$ work same as ordinary
    thermodynamic systems in which $V$ is black hole thermodynamic volume and $P$ is  pressure of the surrounded
  environments. In fact $P$ is equal to inverse of radius square of AdS space (de Sitter space with negative cosmological parameter).
  In a geometrical perspective an AdS (dS) vacuum space is an open hyperbola (closed spherical) with negative (positive) Gaussian spatial
  curvature space time. In other words negative (positive) values cosmological parameter is related to negative (positive)
  values
  vacuum energy density in the AdS (dS) space. There published many papers which one can follow
  in the literature (see for instance
  \cite{Ghaffar5,Ghaffar6,Ghaffar7, Ghaffar8,Ghaffar9, Ghaffar10, Ghaffar11, Ghaffar12, Ghaffar13, Ghaffar14,
   Ghaffar15, Ghaffar16, Ghaffar17} and references therein).
    When one study thermodynamic behavior of the black holes without (with) to use the cosmological constant parameter who
   is apply in fact an ordinary (extended) thermodynamic phase space.
  Studying thermodynamics of black holes is one of the most remarkable and importable subjects to investigate in all gravitational theories and
   physicists have hope to obtain some acceptable proposals about the unknown essential quantum gravity theory
   via studies of black holes thermodynamics.  Moffat presented a particular scalar-tensor-vector gravity (STVG) model \cite{Moffat}
   and he solved its gravitational metric equations to obtain a gravitationally charged spherically symmetric static black hole metric solution.
   His obtained solution is similar more as Reissner Nordstrom electrically charged black hole metric solution
   \cite{Moffat5} where we want to study affects of the black hole charge on its possible thermodynamic phase transition in presence of
    the AdS pressure. We will call this black hole as AdS Schwarzschild STVG black hole in what follows.
Layout of the work is as follows.\\
In section 2 we introduce in summary the AdS Schwarzschild STVG
black hole metric. In section 3 we calculate its thermodynamic
variables such as entropy, temperature, heat capacity, Gibbs free
energy and etc., and try to give a modified Smarr relation. By
plotting diagrams of the thermodynamic variables we  investigate
to analyze possible phase transition of the black hole and
situations where the black hole exhibit with two coexistence
subsystems. Section 3 denotes to the concluding remarks and
outlook of the work.
\section{AdS Schwarzschild STVG black hole}
Let us we start with scalar tensor vector gravity (STVG) model
which was presented at a first time by Moffat \cite{Moffat} as
follows.
\begin{equation}
\label{action}S=S_{Grav}+S_{\phi}+S_S+S_M
\end{equation}
where gravity part of the above action $G_{Grav}$ is defined by
the Einstein Hilbert action with additional cosmological parameter
$\Lambda$\begin{equation} S_{Grav}=\frac{1}{16\pi}\int
d^4x\sqrt{-g}[\frac{1}{G}(R+2\Lambda)],
\end{equation}
the modified massive vector part of the action $S_{\phi}$ is given
by
\begin{equation} S_{\phi}=-\int
d^4x\sqrt{-g}\bigg[\omega\left(\frac{1}{4}B^{\mu\nu}B_{\mu\nu}+V(\phi)\right)\bigg]
\end{equation}
and the scalar part of the action $S_S$ is given by $$ S_S=\int
d^4x\sqrt{-g}\left[\frac{1}{G^3}\left(\frac{1}{2}g^{\mu\nu}\nabla
_{\mu}G\nabla _{\nu}G-V(G) \right)\right]$$$$+\int
dx^4\sqrt{-g}\frac{1}{G}\bigg(\frac{1}{2}g^{\mu\nu}\nabla
_{\mu}\omega\nabla _{\nu}\omega-V(\omega)\bigg)$$
\begin{equation}
+\int
d^4x\sqrt{-g}\left[\frac{1}{\mu^2G}\left(\frac{1}{2}g^{\mu\nu}\nabla
_{\mu}\mu\nabla _{\nu}\mu-V(\mu)\right) \right]
\end{equation}
respectively and $S_M$ denotes to model dependent ordinary matter
source. $g$ is absolute value of the metric determinant and
$R=g_{\mu\nu}R^{\mu\nu}$ is the Ricci scalar. $\phi_{\mu}$ refers
to a massive vector field with mass parameter $\mu$ and self
interaction potential $V(\phi)=-\frac{1}{2}\mu^2\phi
^{\mu}\phi_{\mu}$ and
$B_{\mu\nu}=\partial_\mu\phi_\nu-\partial_\nu\phi_\mu$ is anti
symmetric linear tensor field and $\omega$ is a dimensionless
scalar field with self interaction potential $V(\omega)$. $V(G)$
is self interaction potential of non material scalar field $G(x)$
(namely variable Newton gravity coupling parameter) and $V(\mu)$
denotes to self interaction potential according to the scalar
field $\mu(x)$. $\nabla_{\mu}$ refers to the covariant derivative
for a metric tensor field $g_{\mu\nu}$. Such these alternative
gravity models are called as creative models against GR, because
without the last term $S_M$ these models can create gravity by
self interacting of the fields while in GR itself $S_{Grav},$ the
external matter source $S_{M}$ is necessary to produce the
gravity. In fact effects of the vector field mass $\phi$ dose not
vanishing just at kiloparsec scales from gravitational sources
while so it can be neglected near the black holes solutions of the
model. At the slow varying regime of the Newton's gravity coupling
parameter one can consider $G=G_N(1+\alpha)$ where $G_N$ is the
well known Newton's gravity coupling constant at the Newton and
General relativity approach of the model in which the
dimensionless parameter $\alpha$ comes from alternative
contribution of the above action at the slow varying
 regime of the scalar field $G(x).$ In other words $\alpha$ is usually called as the gravitational charge which for $\alpha=0$
 the STVG gravity returns to GR so we can regard deviation
 of the STVG theory with respect to GR given by $\alpha$ parameter.
For vacuum sector of the action (\ref{action}) the Moffat himself
solved metric field equations and obtained spherically symmetric
static metric vacuum solutions without to use $\Lambda$. He
obtained a singular asymptotically flat metric same as the
Reissner Nordstrom one with particular gravitational charge
$q=\pm\sqrt{G_NG}M$ \cite{Moffat5} as
\begin{equation}\label{mofmet}ds^2=-\big(1-2GM/r+\alpha
G_NGM^2/r^2\big)dt^2\end{equation}$$+\big(1-2GM/r+\alpha
G_NGM^2/r^2\big)^{-1}dr^2+r^2d(d\theta^2+\sin^2\theta
d\varphi^2)$$ for a gauge field $\phi_t(r)\neq0$ with
$\phi_r(r)=\phi_{\theta}(r)=\phi_{\varphi}(r)=0.$ Haydarov et al
\cite{Haydarov} studied affects of the gravitational charge
$\alpha$ on stabilization of orbits of test particles moving on
the space time (\ref{mofmet}) recently. Moffat himself obtained
also another nonsingular asymptotically flat black hole metric
(not shown) by choosing other form for the vector gauge field
$\phi_{\mu}$ in the ref. \cite{Moffat5} where central region of
the obtained metric solution reduces to vacuum de Sitter space
which makes it as nonsingular. By looking at his metric solutions
one can infer that the produced effective cosmological parameter
versus $\alpha$ and the black hole mass $M$ is depended to kind of
the used vector gauge field $\phi_{\mu}$. But we should point that
these asymptotically flat solutions are different with non
asymptotically flat metric solutions which are obtained usually by
regarding a nonzero cosmological parameter $\Lambda\neq0.$ They
approach asymptotically to the vacuum de Sitter $(\Lambda>0)$ or
the vacuum  anti de Sitter $(\Lambda<0)$ spaces. In this sense we
want study thermodynamic behavior of the metric solution
(\ref{mofmet}) with $\Lambda\neq0$
 in this work. Thermodynamic behavior of the modified Schwarzschild black hole
metric (\ref{mofmet}) is studied by Mureika et al \cite{Mureika}
where the
 black hole heat capacity exhibits with change of the sign at the critical mass in presence of the Hawking radiation.
  This means the black hole  can be has a phase transition same as the Reissner Nordsrom black hole itself (i.e. with electric charge) by
  regarding the Hawking temperature and the backreaction
  correction of the interacting quantum fields which is studied by one of us previously \cite{Ghaffar5}.
  In general the first law of (the asymptotically flat) black hole thermodynamics is usually written as
  \begin{equation}\label{firstlaw0} dM=TdS+\Omega dJ+\Phi dQ
   \end{equation}
  where $T=\frac{\kappa}{4\pi}$ is the Hawking temperature of the black hole
    ($\kappa$ is the surface gravity), $S=\frac{A}{4}$ is  the Bekenstein Hawking
    entropy of the  black hole
     ($A$ is the event horizon surface area), $\Omega$ is the angular velocity, $J$ is the angular
    momentum, $\Phi$ is the electrostatic potential difference
    between infinity and the horizon, $Q$ is the electric
    charge and $M$ is the black hole mass. Usually $M$ is considered to be internal energy of the black
    hole $U$ in the ordinary thermodynamic sense, but it was suggested in \cite{Kastor} that it is more correctly interpreted
    as the enthalpy $M=H=PV+U$ of the black holes in presence of the cosmological parameter.
    In this sense, the first law of the black hole thermodynamics
    (\ref{firstlaw0}) should be extended with variation of the cosmological term $VdP$ as follows \cite{Dolan}.
    \begin{equation}\label{firstlawex} dM=TdS+VdP+\Omega dJ+\Phi dQ
   \end{equation}
  where $P=\frac{-\Lambda}{8\pi}$
  is pressure of the AdS vacuum space  and it is related to radius of the AdS space as
   $\ell_{AdS}=\sqrt{\frac{-3}{\Lambda}}$ for $\Lambda<0.$
   One can infer that the AdS black hole thermodynamic volume can be calculated from the above extended first law of the
   thermodynamics of the AdS black holes
 as $V=\frac{\partial M}{\partial P}\big|_{S,J,Q}.$ In this sense $V$ is interpreted as conjugate thermodynamic variable for the pressure
 $P$ of the AdS space. In other words  $V$ is a finite, effective volume for the region outside the AdS black hole horizon,
 which can also be interpreted as minus the volume excluded from a spatial slice by the black hole horizon.
 In fact the black hole solutions with a non-vanishing cosmological constant $\Lambda$
   have received considerable recent attention because of two reasons as follows: This is due both to the role they play in the
   phenomenology of the AdS/CFT
   correspondence \cite{Malda}, \cite{Gubser}, \cite{Witten} which associates the cosmological constant with the rank of the gauge
   group originally
and also, of course, to the observational data suggesting that the
universe may have a small positive value of $\Lambda$
\cite{spergel}.  In four dimensions, the extension of the
Kerr-Newman family of solutions to non-zero $\Lambda$ term was
found many years ago by Carter \cite{carter}. In fact many authors
are investigating to bring our understanding of certain properties
of AdS black holes more closely in parallel with well known
results in the asymptotically flat case. Dolan showed in ref.
\cite{Dolan} that the cosmological parameter raises efficiency of
Penrose process in the AdS black hole with respect to case where
$\Lambda=0.$ Because of importance of $\Lambda$ term which we
introduced here and also as an extension of our previous work
\cite{Ghaffar5} we like to study thermodynamic behavior of the AdS
gravitationally charged Schwarzschild STVG black hole which is
given by the subsequent line element.
\begin{equation}
ds^2=-f(r)dt^2+f(r)^{-1}dr^2+r^2d\Omega ^2
\end{equation}
where the metric potential $f(r)$ is
\begin{equation}\label{met}
f(r)=1-\frac{2(1+\alpha)M}{r}+\frac{\alpha
(1+\alpha)M^2}{r^2}+\frac{8\pi}{3}Pr^2
\end{equation}
for which we substitute $P=\frac{-\Lambda}{8\pi},$
$G=G_N(1+\alpha)$ and $G_N=1.$ To study thermodynamic behavior of
this AdS Schwarzschild STVG black hole, let us to use an
equipotential surface $f(r,M,P, \alpha)=constant$ to obtain first
law of thermodynamics for this black hole by varying this
equipotential surface with respect to the variables $r,M,\alpha,
P$ and by setting $df(r,M,P, \alpha)=0$ as follows.
\begin{equation}\label{first}dM=TdS+VdP+\Phi_{\alpha}d\alpha\end{equation}
where the event horizon hypersurface $r=r_h$ is determined by
setting  $f(r_h,\\M,P, \alpha)=0$ as follows.
\begin{equation}\label{event}1-\frac{2(1+\alpha)M}{r_h}+\frac{\alpha(1+\alpha)M^2}{r_h^2}+\frac{8\pi r_h^2P}{3}=0.\end{equation}
 In the equation (\ref{first}) the Hawking
temperature is given versus the surface gravity of the event
horizon as
\begin{equation} \label{haw}
T=\frac{1}{4\pi}\frac{df}{dr}\bigg|_{r_h,
M,P,\alpha}=\frac{4r_hP}{3}+\frac{(1+\alpha)M}{2\pi
r_h^2}\bigg(1-\frac{\alpha M}{r_h}\bigg)\end{equation} and
\begin{equation}\label{ent}dS=\frac{2\pi r_hdr_h}{(1+\alpha)\big(1-\frac{\alpha M}{r_h}\big)}
\end{equation}is the entropy difference,
and thermodynamic volume
\begin{equation}\label{Vol}V=\frac{\frac{4\pi
r_h^3}{3}}{(1+\alpha)\big(1-\frac{\alpha M}{r_h}\big)}
\end{equation}
is conjugate variable of the AdS space pressure $P$ and
\begin{equation}\label{Pot}\Phi_{\alpha}=-\frac{M}{2(1+\alpha)}\bigg(\frac{2r_h-(1+2\alpha)M}{r_h-\alpha
M}\bigg)\end{equation} is the conjugate variable for dimensionless
gravitational charge $\alpha.$  One can see that the entropy
deference (\ref{ent}) reads to the Bekenstein-Hawking
 entropy $S=\frac{A}{4}=\pi r_h^2$ for $\alpha=0$ and the corresponding thermodynamic volume (\ref{Vol})
 reduces to the geometric volume of the black hole $V=\frac{4\pi r_h^3}{3}.$ While for $\alpha\neq0$ and $M=H=constant$
 the integration of the entropy difference
 (\ref{ent}) leads to the following equation containing a logarithmic term.
 \begin{equation}\label{entropy} S=\int_{0}^{r_h}\frac{2\pi r_hdr_h}{(1+\alpha)\big(1-\frac{\alpha M}{r_h}\big)}
 =\frac{2\pi}{(1+\alpha)}\bigg\{r_h^2+\alpha M r_h+\alpha^2 M^2\ln\bigg|1-\frac{r_h}{\alpha M
 }\bigg|\bigg\}.\end{equation} In fact this logarithmic term originates from quantum aspect which can be followed via
\cite{Don} and references therein. To obtain a relationship
between the thermodynamic variables of this black hole to be
independent of the geometric parameter we can do as follows:
  By substituting
$(1-\alpha M/r_h)$ from (\ref{Vol}) into the equation (\ref{haw})
we obtain
\begin{equation}\label{rh}r_h=\frac{3TV}{2(M+2PV)}.\end{equation}
Next we substitute the latter equation into the conjugate
potential (\ref{Pot}) to obtain the following identity between
thermodynamic variables of the black hole.
 \begin{equation}\label{Smarr}\Phi_{\alpha}=\frac{-M}{4(1+\alpha)}\bigg[\frac{(1+2\alpha)M(M+2PV)-3TV}{2\alpha M(M+2PV)-3TV}\bigg].
 \end{equation} This is differ with the well known Smarr relation \cite{Smarr} obtained from dimensional approach as $M=2TS+\alpha\Phi_\alpha$
  \footnote{To obtain $M=2TS+\alpha\Phi_\alpha$ we set $Q=\alpha M$ and $\Phi_Q= \frac{\Phi_{\alpha}}{M}$ and substitute
  them into the Smarr relation of AdS Reissner-Nordstrom black hole which was given by $M=2TS+\Phi_QQ$   in ref.
  \cite{Shi}.}
  for the metric equation (\ref{met}).
   However there is do not
  worry because authors of the ref. \cite{Shi} showed also that there are some black hole solutions which do not obey completely the
  Smarr relation of the black holes. In this sense we can claim the equation (\ref{Smarr}) is in fact modified Smarr relation for
   the AdS Schwarzschild STVG black hole under consideration. The equation (\ref{Smarr}) is a hypersurface
  $F(\Phi_\alpha,\alpha,P,V,T,M)=0$ defined in a 6-dimensional
  phase space. It is useful to obtain inflection point of this
  hypersurface by calculating \begin{equation}\label{inflect}\frac{\partial \Phi_{\alpha}}{\partial\alpha}\bigg|_{P,V,T}=0,~~
  ~\frac{\partial^2\Phi_\alpha}{\partial\alpha^2}\bigg|_{P,V,T}=0\end{equation}
which reads to 3 different critical points in 6 dimensional phase
space $\{\Phi_\alpha,\alpha,\\ P,V,T,M\}$ as follows.
\begin{equation}I:~~~~~~~~~\alpha\to\infty,~~~\Phi_\alpha=0,~~~~M(2PV+M)=0\end{equation}
\begin{equation}II:~~~~\alpha=-1,~~~~\Phi_\alpha\to-\infty,~~~~2M^2+4MPV+3TV=0
\end{equation}
and
\begin{equation}III:~~~~\alpha=-\frac{3}{4},~~~\Phi_{\alpha}=4M,~~~~M^2+2MPV+3TV=0.\end{equation}
In the following we will see that none of the above values for
$\alpha=-1,-\frac{3}{4}$ and or $\alpha\to\infty$ do not describe
physical situations and so possibility of phase transition is done
just for $\alpha=1.$\\
 We proceed now to obtain possible critical points of this black
hole system as follows. By regarding the black hole evaporation in
presence of the Hawking radiation and beak-reaction corrections of
interacting quantum fields the mass of the black hole is lost
\cite{Ghaffar2}, \cite{Ghaffar3}, \cite{Ghaffar5} and so
thermodynamics study of black holes should be done with variable
enthalpy $M$. This leads us to obtain an equation of state for the
AdS Schwarzschild STVG black hole (\ref{met}) which is independent
of the mass $M.$ To do so we first obtain the possible critical
points which occur through the following conditions for the
inflection points.
\begin{equation}\label{critical}\frac{\partial T}{\partial r_h}\bigg|_{P,M,\alpha}=0,~~~\frac{\partial^2 T}{\partial r_h^2}\bigg|_{P,M,\alpha}=0
\end{equation} which by substituting the Hawking temperature
(\ref{haw}) reads
\begin{equation}\label{critical1}r_c=2\alpha M,~~~P_c=\frac{3(1+\alpha)}{128\pi\alpha^3M^2},~~~T_c=\frac{(1+\alpha)}{8\pi\alpha^2M}.\end{equation}
  By substituting (\ref{critical1}) into
the thermodynamic volume (\ref{Vol}) we will have for critical
thermodynamic volume
\begin{equation}\label{Volc}V_c=\frac{64\pi
\alpha^3M^3}{3(1+\alpha)}.\end{equation} The critical horizon
radius $r_c$ and the pressure $P_c$ given by (\ref{critical1})
must be obey the equation of horizon (\ref{event}) for which
 we obtain the following condition.
\begin{equation}\alpha=1.\end{equation} This corresponds to extremis condition on the Reissner Nordstrom charged black hole which is called as
Lukewarm black hole \cite{Ghaffar4} where the electric charge of
the black hole is equal to the mass and so the black hole
temperature vanishes \cite{Ghaffar5}. To make a dimensionless
equation of state we defined
\begin{equation}\label{dim}t=\frac{T}{T_c},~~~~p=\frac{P}{P_c},~~~v=\frac{1}{2}\bigg(\frac{r_h}{r_c}\bigg)\end{equation}
and substitute into the temperature equation (\ref{haw}) so that
\begin{equation}\label{EQt}t=pv+\frac{1}{4v^2}-\frac{1}{16v^3}.\end{equation} The above equation of state
 reduces to the well known ideal
gas equation of state for large black holes $v\to\infty$ where we
must be call $v$ as specific volume of this black hole. In this
sense the obtained critical point will be
\begin{equation}\label{crit}
(t_c,p_c,v_c)=\bigg(1,1,\frac{1}{2}\bigg)\end{equation} This
critical point is applicable for every AdS Schwarzschild STVG
black hole with arbitrary mass. Now that we have found a
mass-independent equation of state we investigate its possible
phase transition for different processes of isotherm and isobaric
namely for processes at constant temperature and at constant
pressure respectively. Substituting (\ref{critical1}) and
(\ref{dim}) into the entropy (\ref{entropy}) and the potential
(\ref{Pot}) one can obtain dimensionless forms respectively for
them as follows.
\begin{equation}s=4v^2+v+\frac{1}{4}\ln|1-4v|\end{equation}
and
\begin{equation}\phi=\frac{8v-3}{4v-1}\end{equation}
where we defined $s=\frac{S}{4\pi M^2}$ and
$\phi=\frac{\Phi_\alpha}{\Phi_c}$ in which critical potential is
$\Phi_c=\Phi_\alpha(r_c)=-\frac{M}{4}.$ Substituting the above
dimensionless thermodynamic variables into the Gibbs free energy
$G=M-TS$ one can obtain its dimensionless form as follows.
\begin{equation}g=-\frac{1}{2}+3ts\end{equation}
in which we defined $g=\frac{G}{G_c}$ and $G_c=-\frac{M}{2}$ is
critical value of the Gibbs free energy.
\section{Thermodynamic phase transition}
 To study possible phase transition in $p-v$ plane which is applicable for isotherm processes
we can rewrite the equation of state (\ref{EQt}) versus the
pressure as follows.
\begin{equation}\label{EQp}p=\frac{t}{v}-\frac{1}{4v^3}+\frac{1}{16v^4}.\end{equation}
We plot diagram of the above equation of state for constant
temperatures below and upper of the critical temperature $t_c=1$
and also corresponding Gibbs free energy in figure 1.
 The pressure diagram in the figure 1 shows the black hole is made from two subsystems for $t<t_c$ but it has a single state for $t>t_c.$
 Furthermore in the Gibbs free energy diagram we can infer that the system is maintain stable for negative values of the Gibbs free energy.
 Physical interpretation of this diagrams is a phase transition from small to large black holes because the Gibbs energy diverges to
 positive infinite value in limits $v\to0$.
 At all system will be unstable where the Gibbs free energy has positive values.
  Also one can see that the pressure diagram in figure 1 is same as
 one which is happened for a Van der Waals fluid in ordinary
 thermodynamics (see \cite{Ghaffar14}). In figure 2 we plot temperature versus the
 entropy, the Gibbs free energy versus the specific volume and
 the temperature for isobaric processes. Diagram of the temperature shows  for a given temperature
  one can obtain two different values for the entropy at $p<p_c$  which means that the black hole system under consideration
  is included with two subsystems (two different phases).
Variation of the Gibbs free energy versus the specific volume at
constant pressure in figure 2 shows that for $p<p_c$ the system
exhibit with the small to large black holes phase transition.
Looking at this diagram one can infer that at constant pressure
for large black holes $v>v_c(=0.5)$ the Gibbs free energy takes
negative values which means the system become stable. In figure 3
we plotted heat capacity of the system at constant pressure.
Changing of the sign of the heat capacity at constant pressure for
$(t,v)\leq(t_c,v_c)$ means that there is happened a phase
transition for the system. When heat capacity has positive
(negative) values the system is called as diathermal or heat
absorber (exothermic or heat repellent). Variation of the heat
capacity  diagram versus the temperature and the specific volume
and its divergency shows that the phase transition is same as one
which is called as second kind  phase transition in ordinary
thermodynamic processes. It may be useful we study behavior of the
compressibility coefficient $\kappa$ and coefficient of the volume
expansion $\beta$ near the critical point. In ordinary
thermodynamics they are called as follows.
\begin{equation}\kappa=-\frac{1}{v}\bigg(\frac{\partial v}{\partial p}\bigg)_t,~~~~\beta=\frac{1}{v}\bigg(\frac{\partial v}{\partial t}\bigg)_p
\end{equation} which corresponding diagrams are plotted in figure 4.
These diagrams show that $\beta$ and $\kappa$ diverge to infinity
at the critical point (\ref{crit}). Looking at the compressibility
coefficient diagram in figure 4 one can see a divergency at small
scale black holes which means that for the black hole exhibit with
a phase transition. Diagram of the volumetric expansion
coefficient versus the temperature shows that this coefficient
decreases by raising the temperature at constant pressure. By
looking at the diagram of the volumetric expansion coefficient vs
the volume one can see that by rasing the specific volume and
reducing it to belove the critical volume the volumetric expansion
coefficient decreases to zero value. In short one can obtain same
interpretation for the system from $\beta$ diagram in figure 4.
\section{Conclusion}
In this paper we studied physical effects of a dynamical vector
field on thermodynamic phase transition of modified AdS
Schwarzschild black hole. In short dynamical vector field creates
a dimensionless $\alpha$ parameter which behaves as charge
parameter for the black hole and so the metric is same as the
Reissner Nordstrom one. We obtained an generalized Smarr relation
for the black hole. Mathematical calculations show small to large
black holes phase transition which is happened at below the
critical point in phase space when the gravitational charge
$\alpha$ of the black hole is equal to the mass $M$. Motivation of
use of AdS space pressure in studying of the black hole
thermodynamics shows that this black hole behaves as ideal gas in
its large scale but at smale scales transit to an imperfect Van
der Waals fluid phase. In this sense without to use AdS pressure
the  phase transition does not happened for the black hole. The
results of this paper are from studies on the behavior of the
Gibbs free energy, heat capacity, volumetric expansion coefficient
and compressibility coefficient on the PVT phases space. At the
future we like to investigate that does occur the other
thermodynamic phenomena for instance affects of holographic
entanglement entropy on thermalization of this black hole? or it
has the Joule-Thomson effect.

\begin{figure}
\label{1} \centering
    \includegraphics[width=4.4cm]{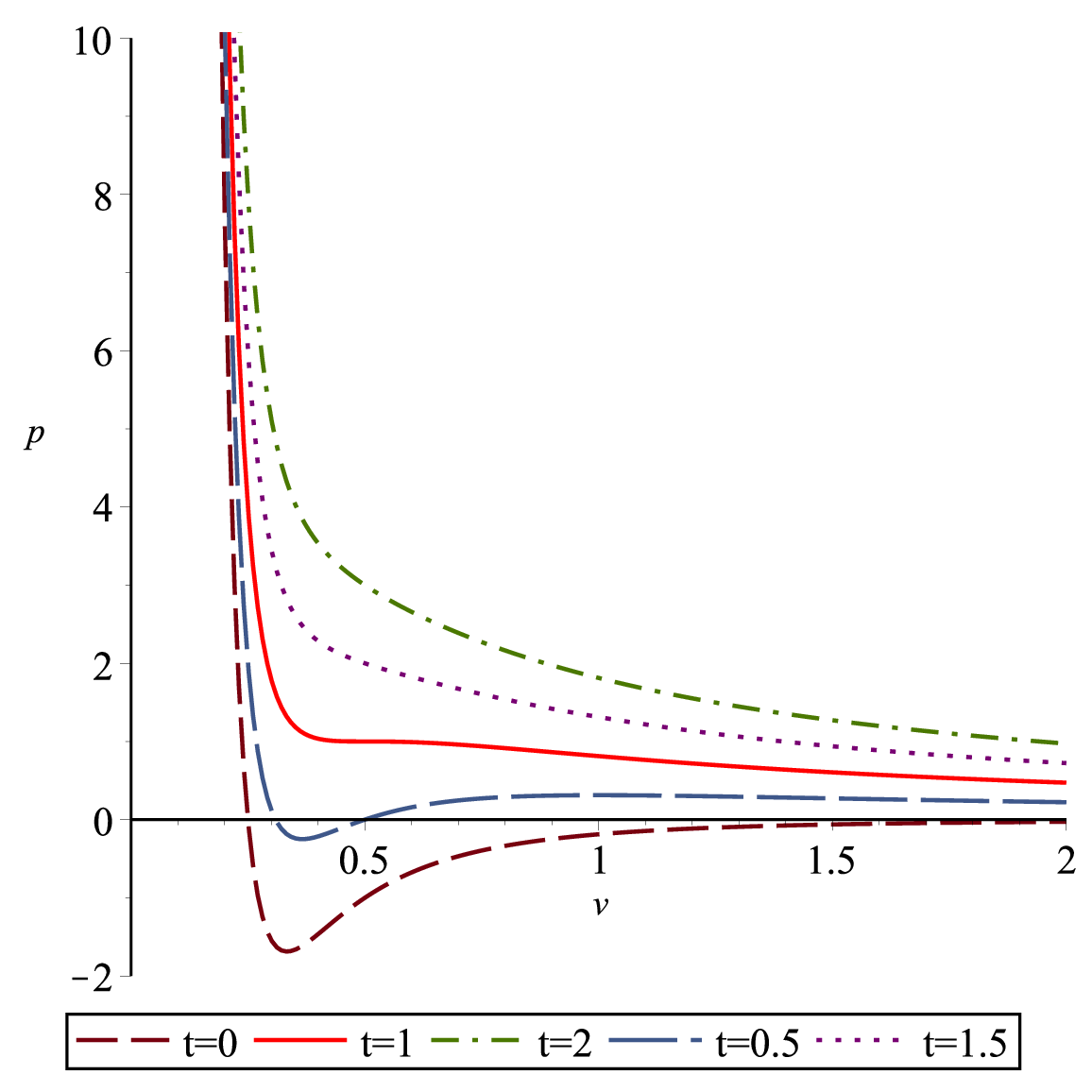}
      \includegraphics[width=4.4cm]{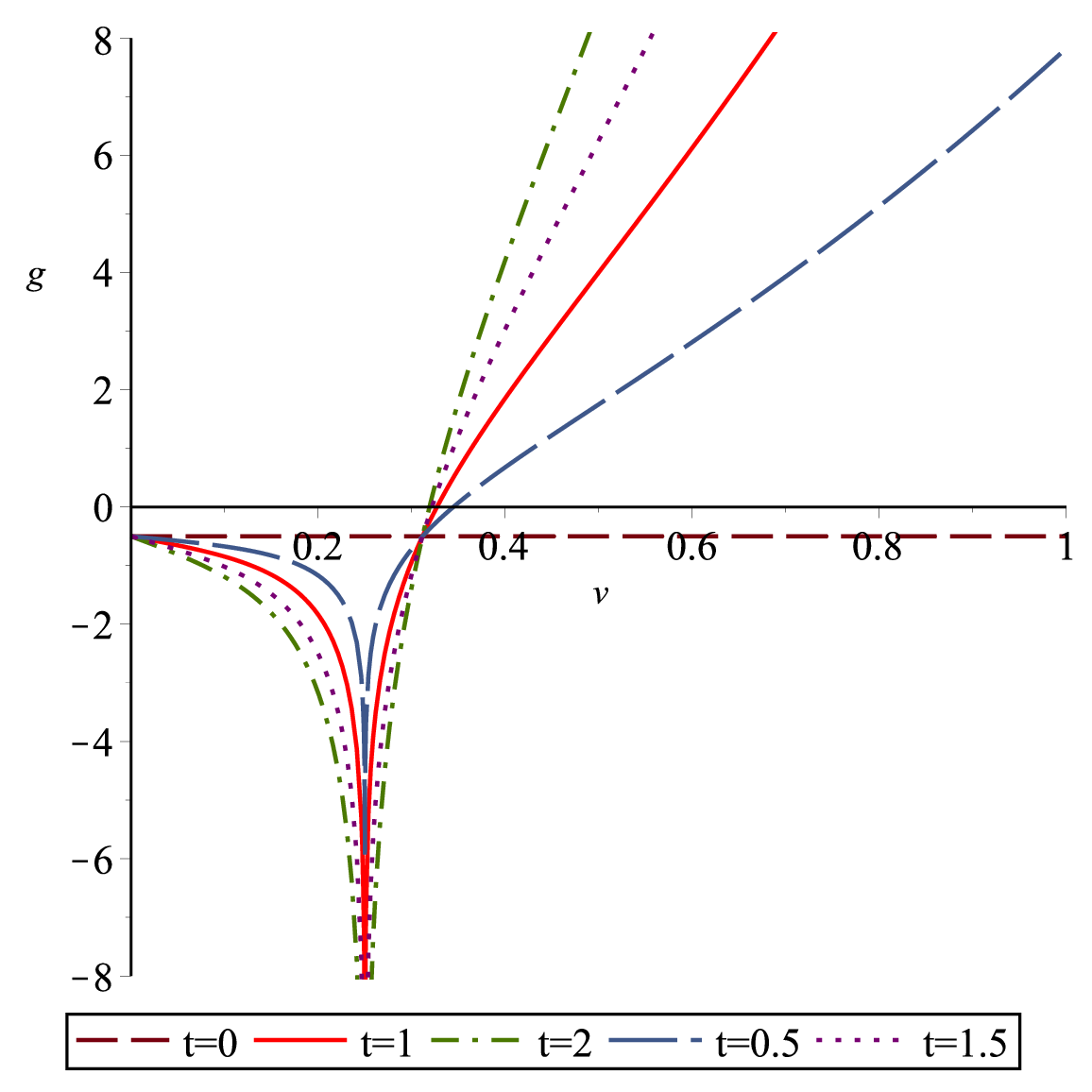}
 \includegraphics[width=4.4cm]{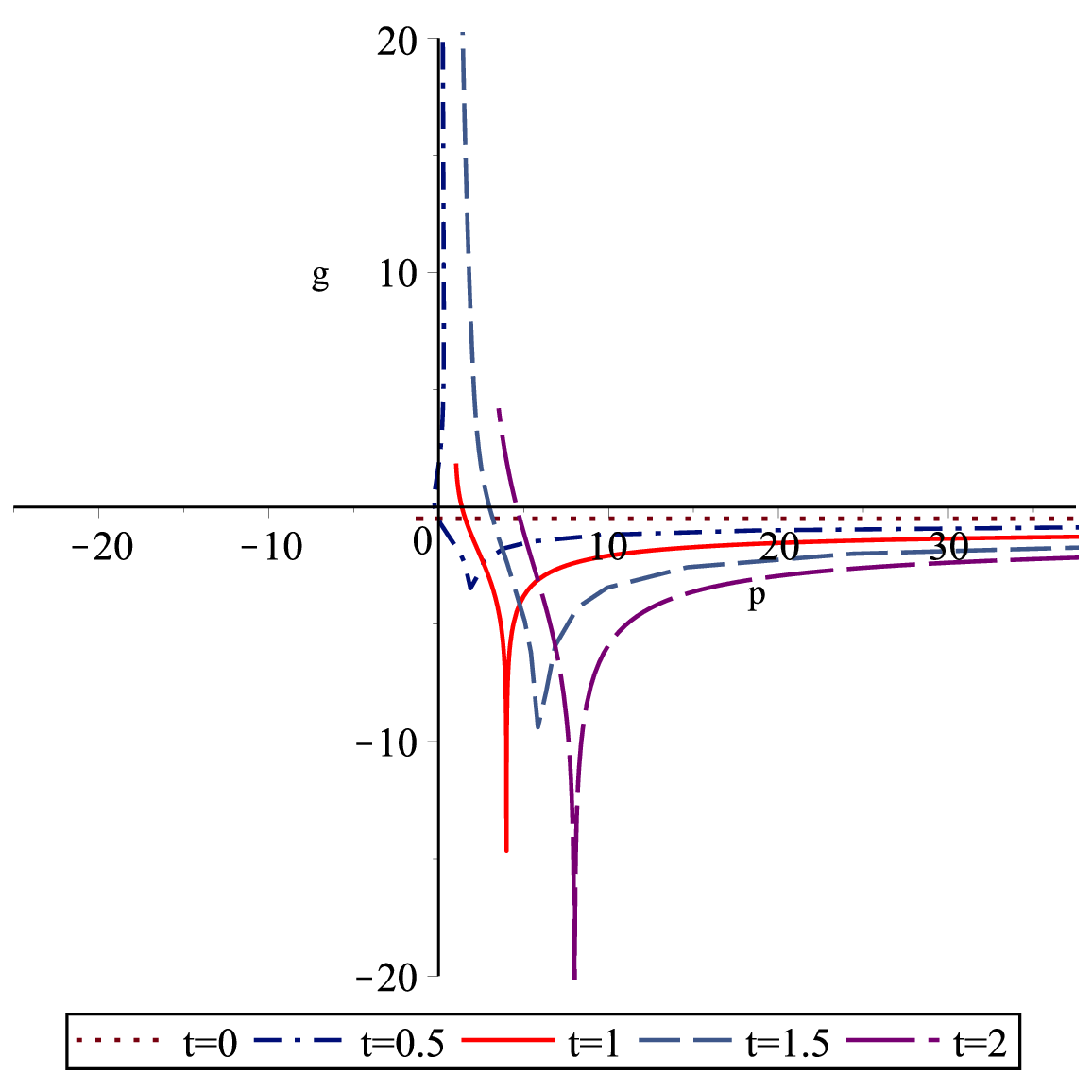}
 \caption{\footnotesize{Diagrams of the pressure $p$ and the Gibbs free energy $g$ are plotted versus the specific volume
 for isotherm processes $t=constnat.$}}
\end{figure}
\begin{figure}
\label{1} \centering
  \includegraphics[width=4.4cm]{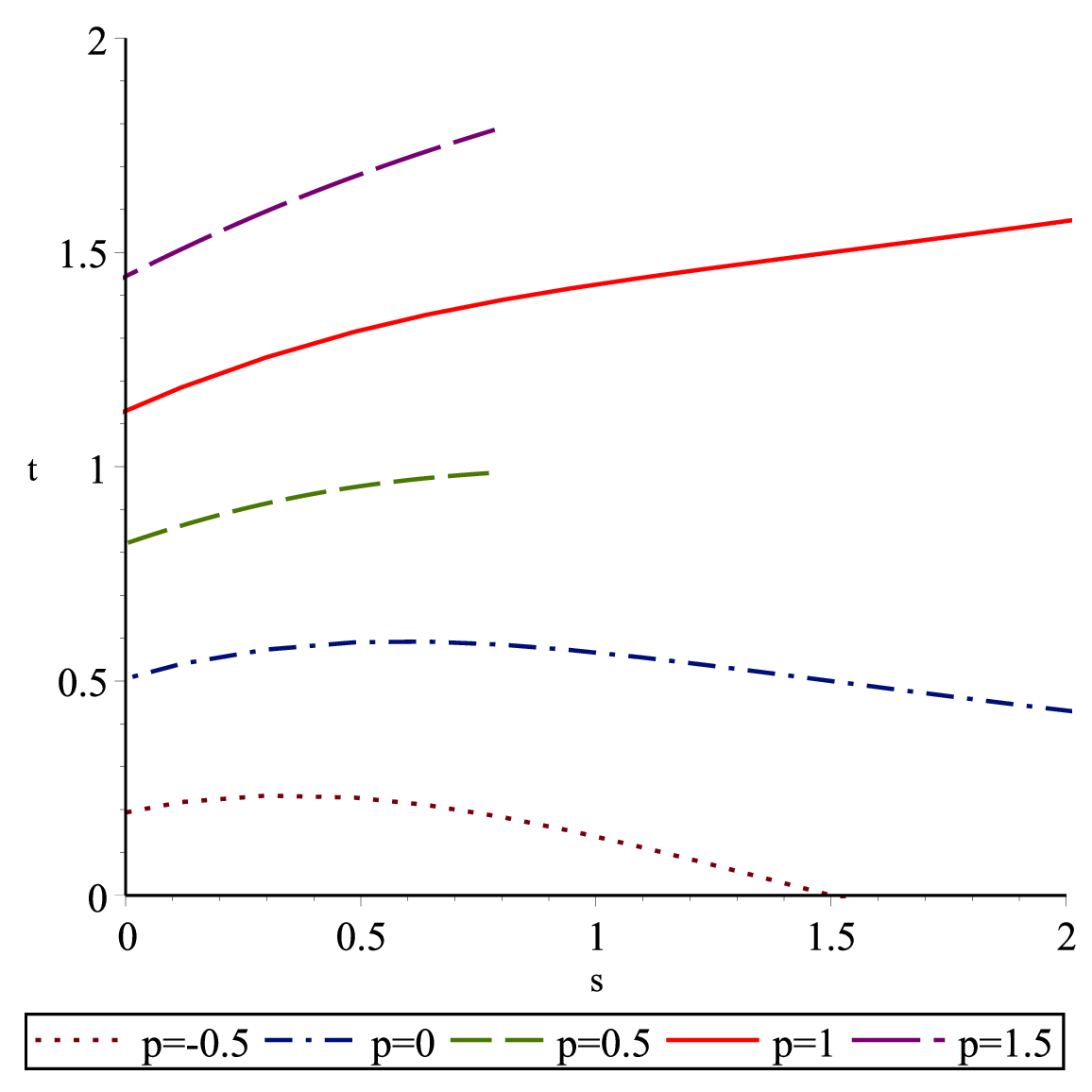}
      \includegraphics[width=4.4cm]{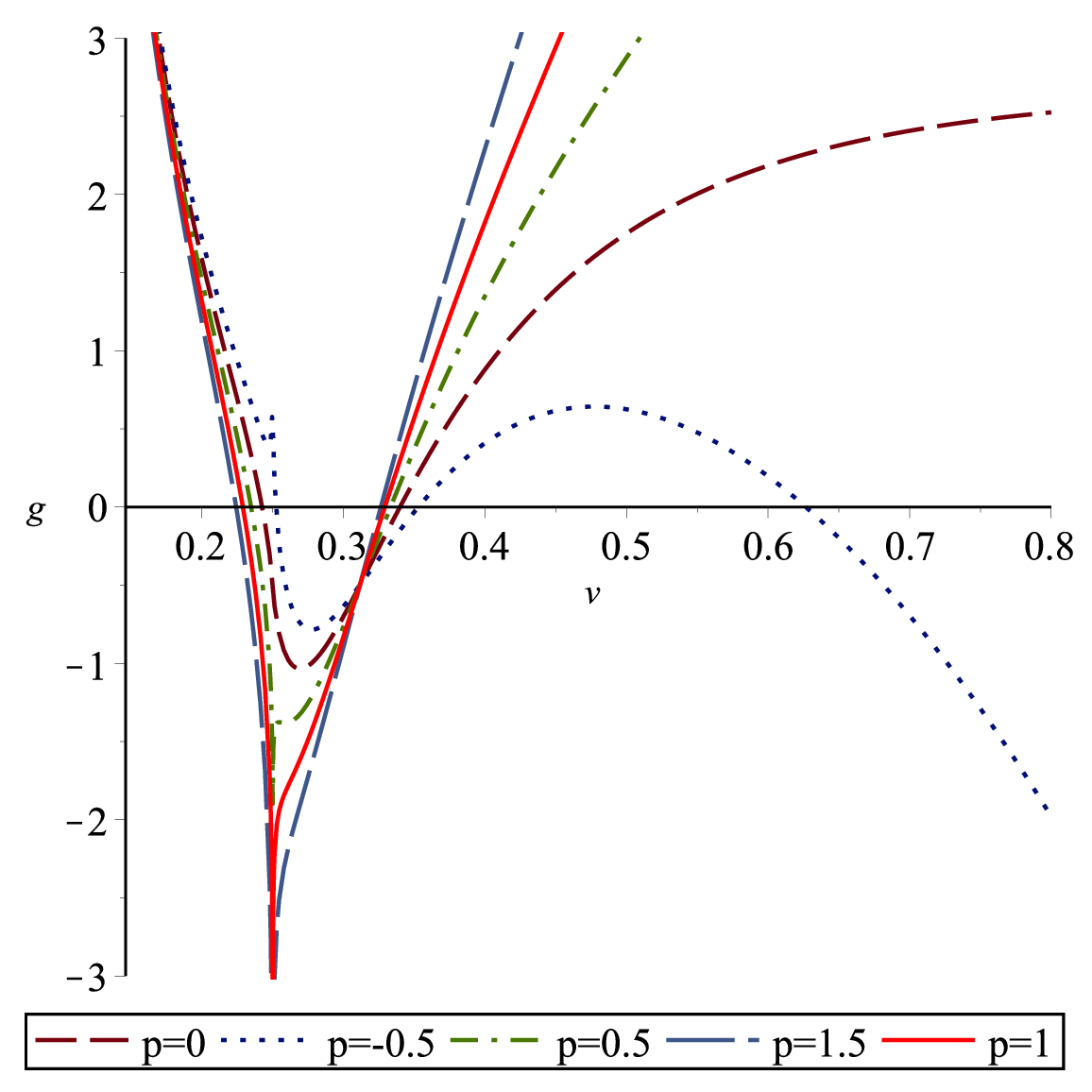}
 \includegraphics[width=4.4cm]{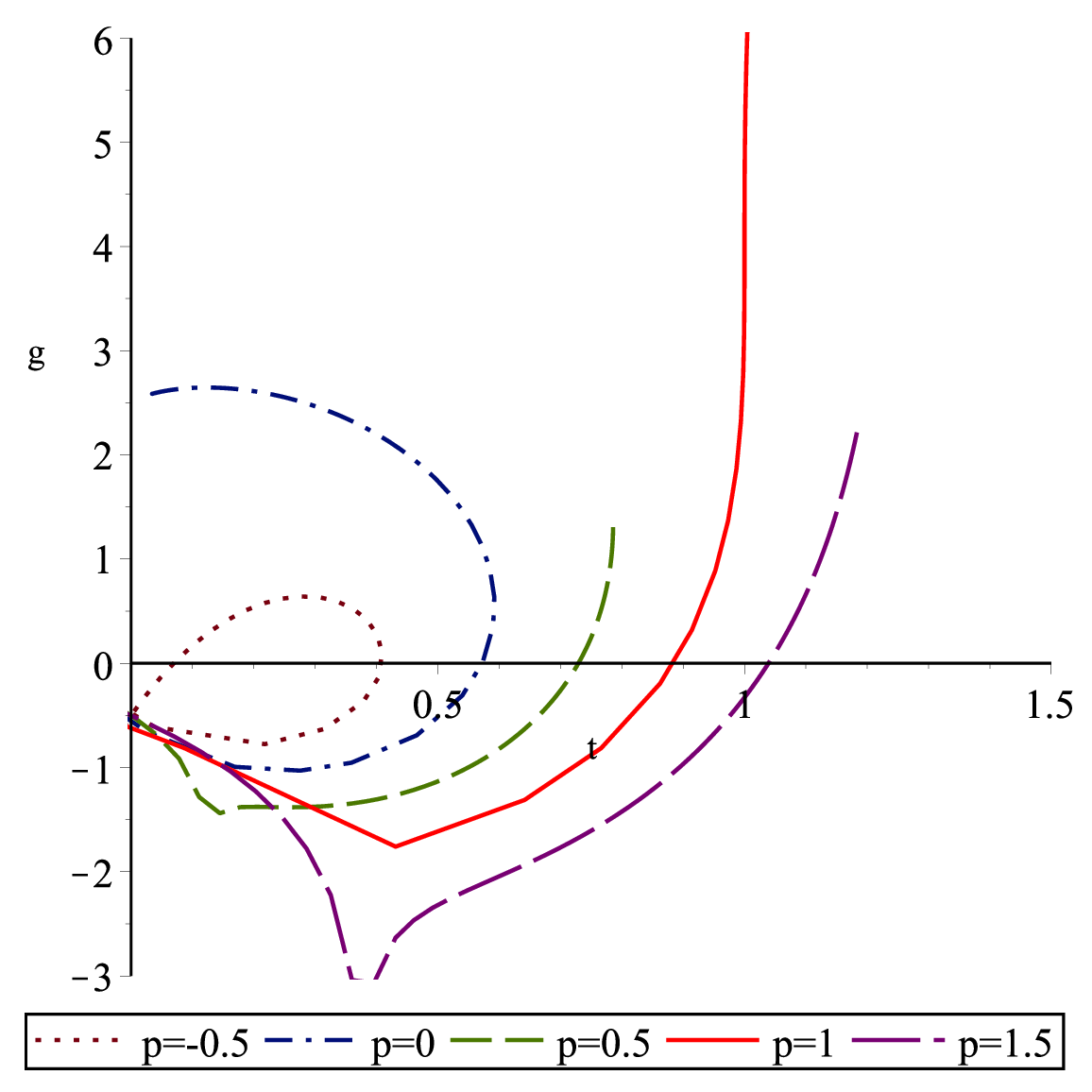}
 \caption{\footnotesize{Diagrams for  the temperature is plotted vs
the entropy at constant pressure and the Gibbs energy is plotted
vs temperature and specific volume  for isobaric processes.}}
\end{figure}
\begin{figure}
\label{1} \centering
  \includegraphics[width=4.4cm]{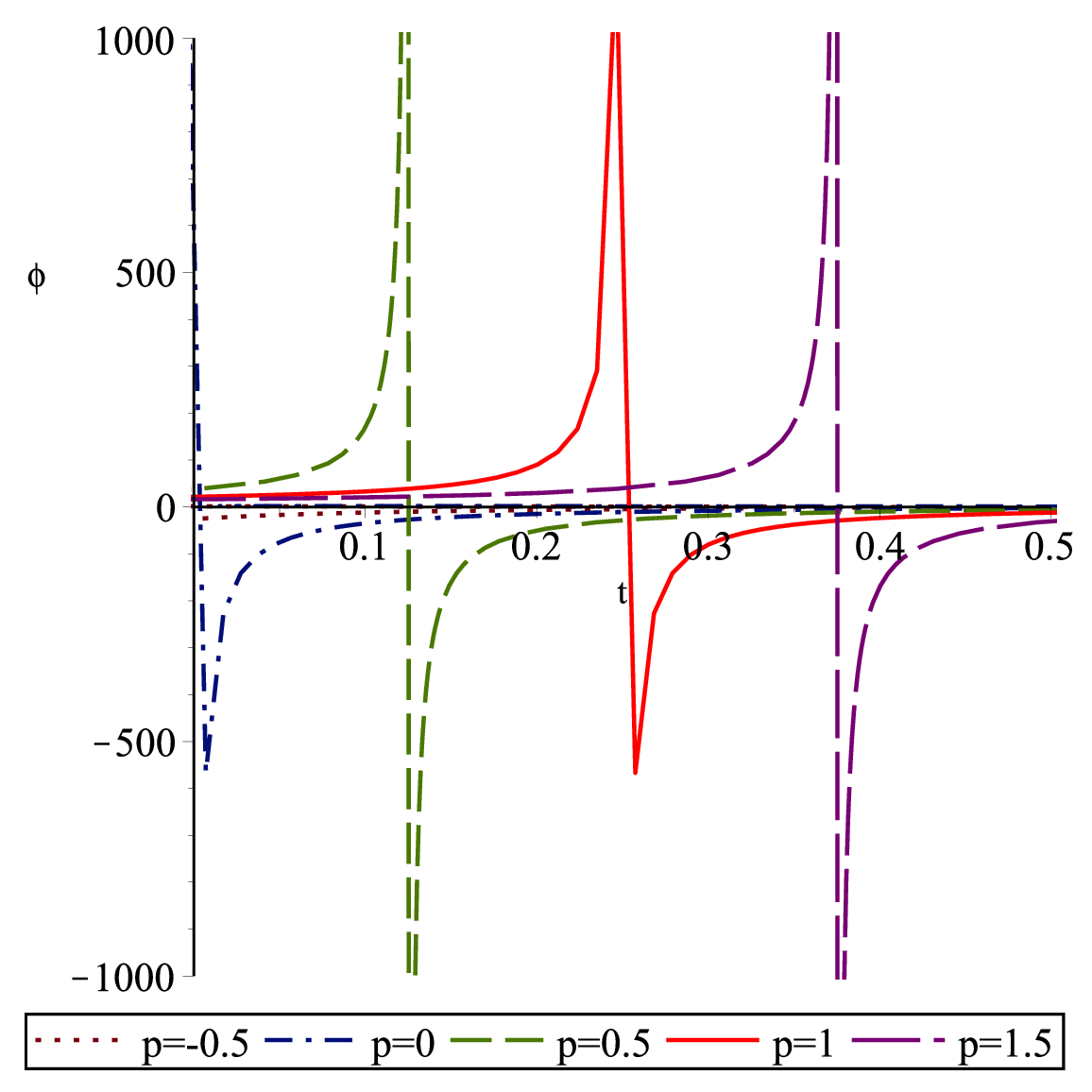}
      \includegraphics[width=4.42cm]{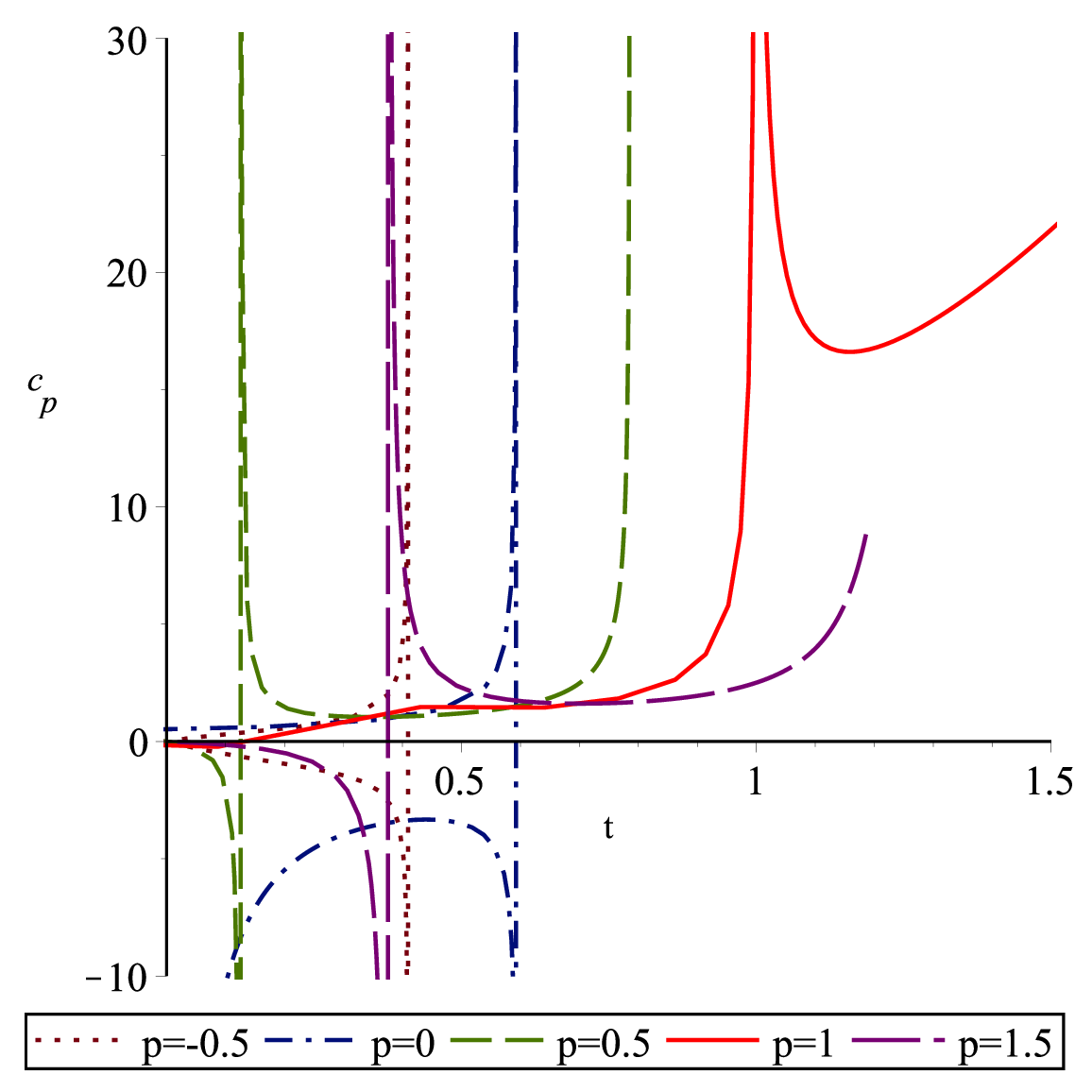}
 \includegraphics[width=4.4cm]{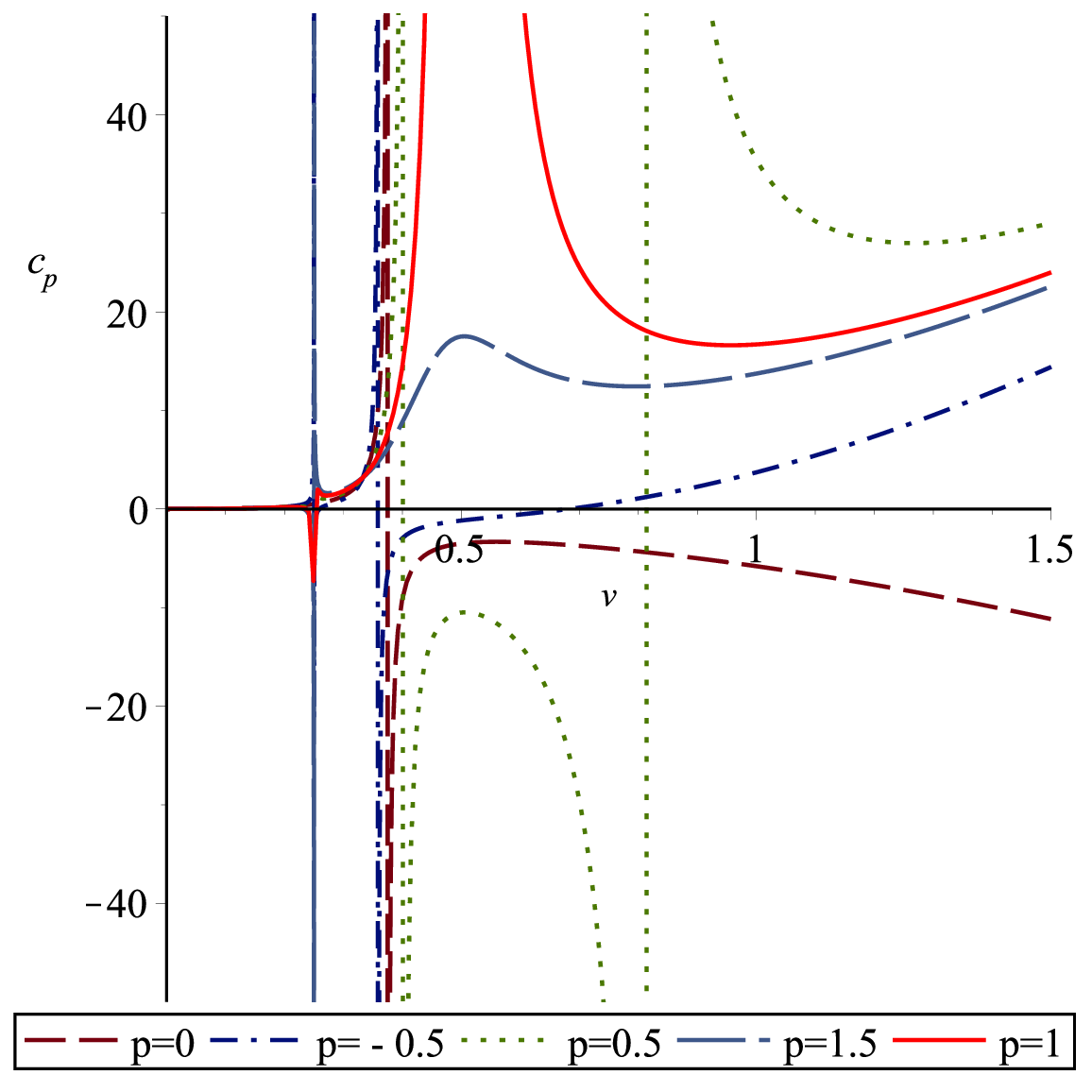}
\caption{\footnotesize{Diagrams for the potential is plotted vs
the temperature at constant pressure and heat capacity is plotted
vs the temperature and the specific volume at constant pressure.
}}
\end{figure}
\begin{figure}
\label{1} \centering
  \includegraphics[width=5cm]{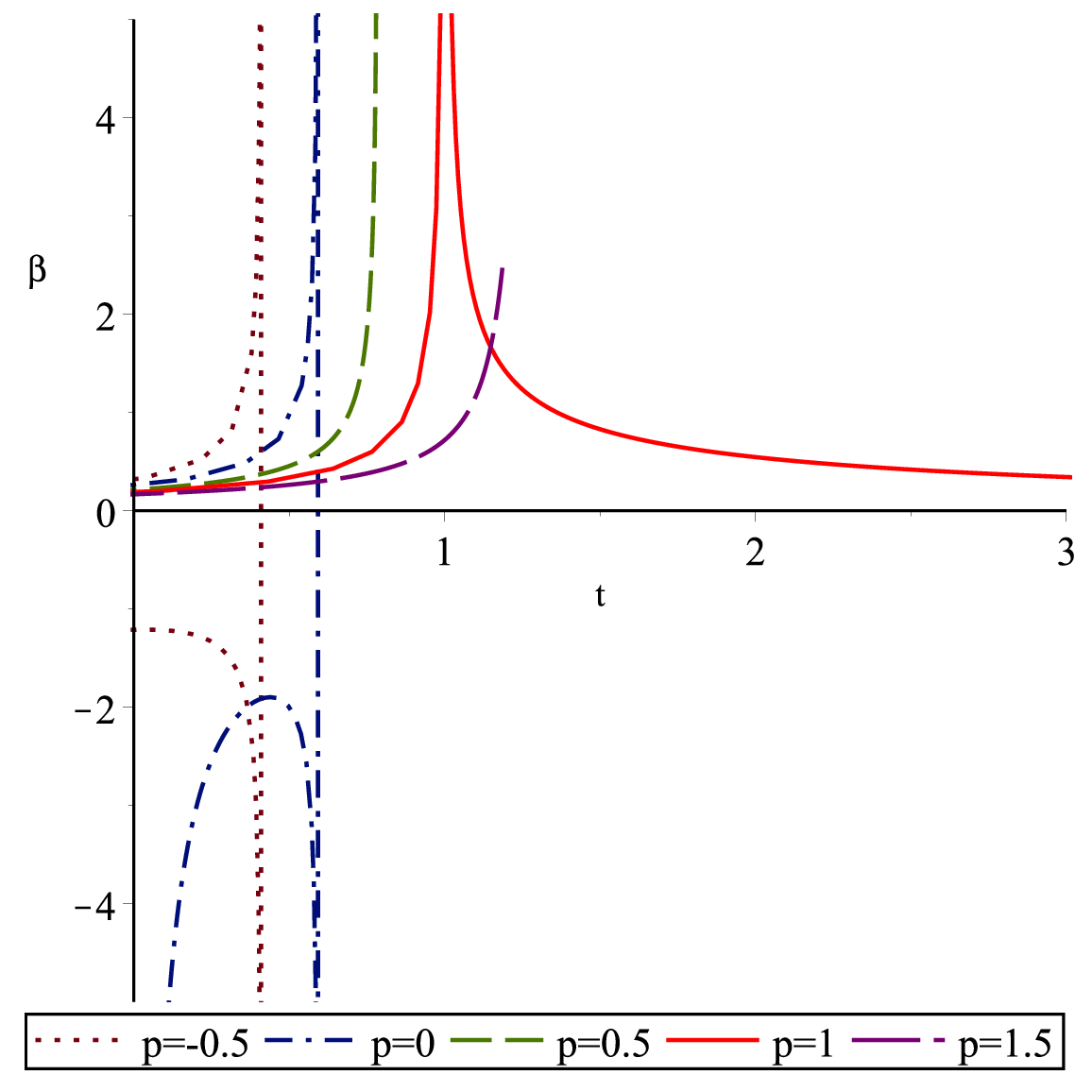}
      \includegraphics[width=5cm]{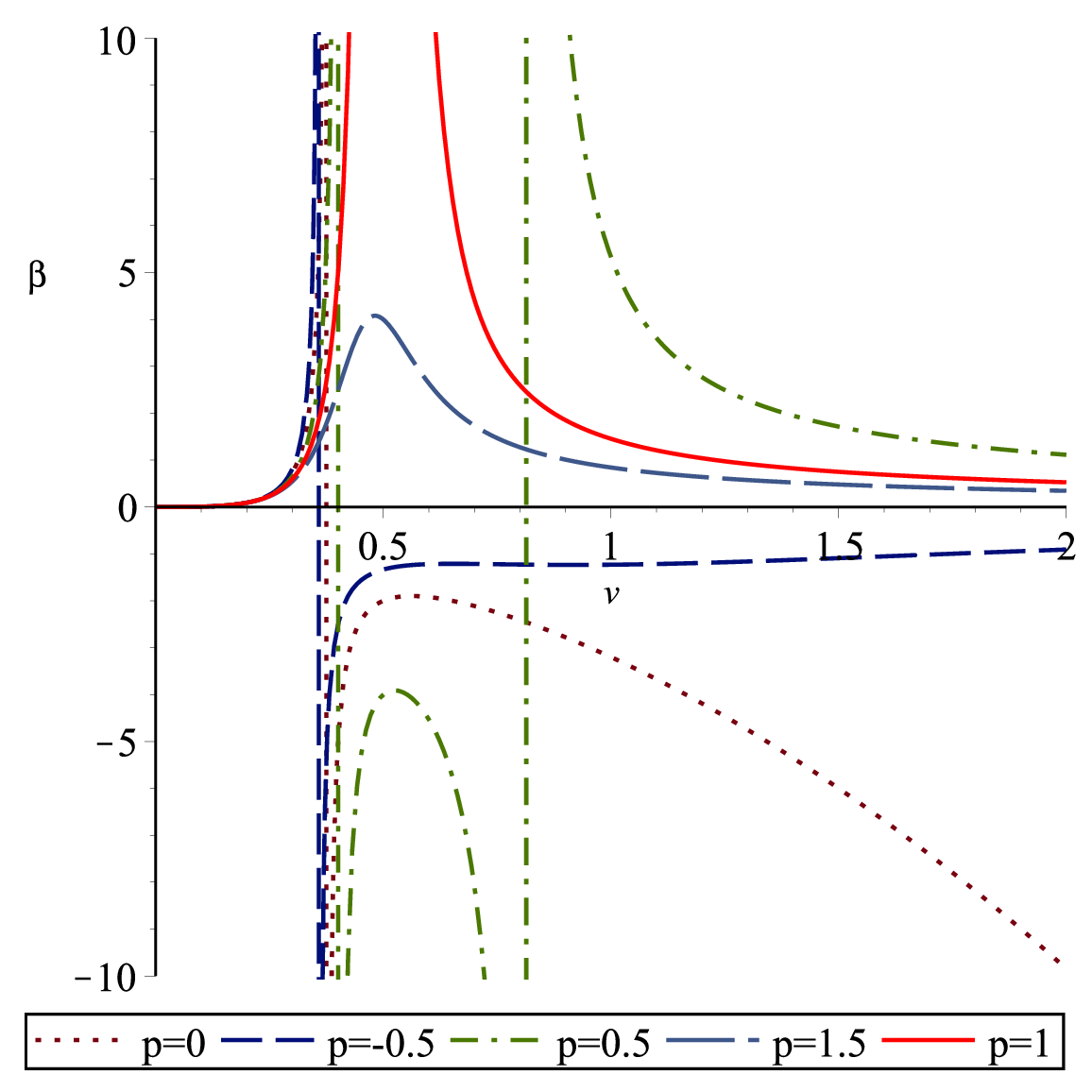}
 \includegraphics[width=5cm]{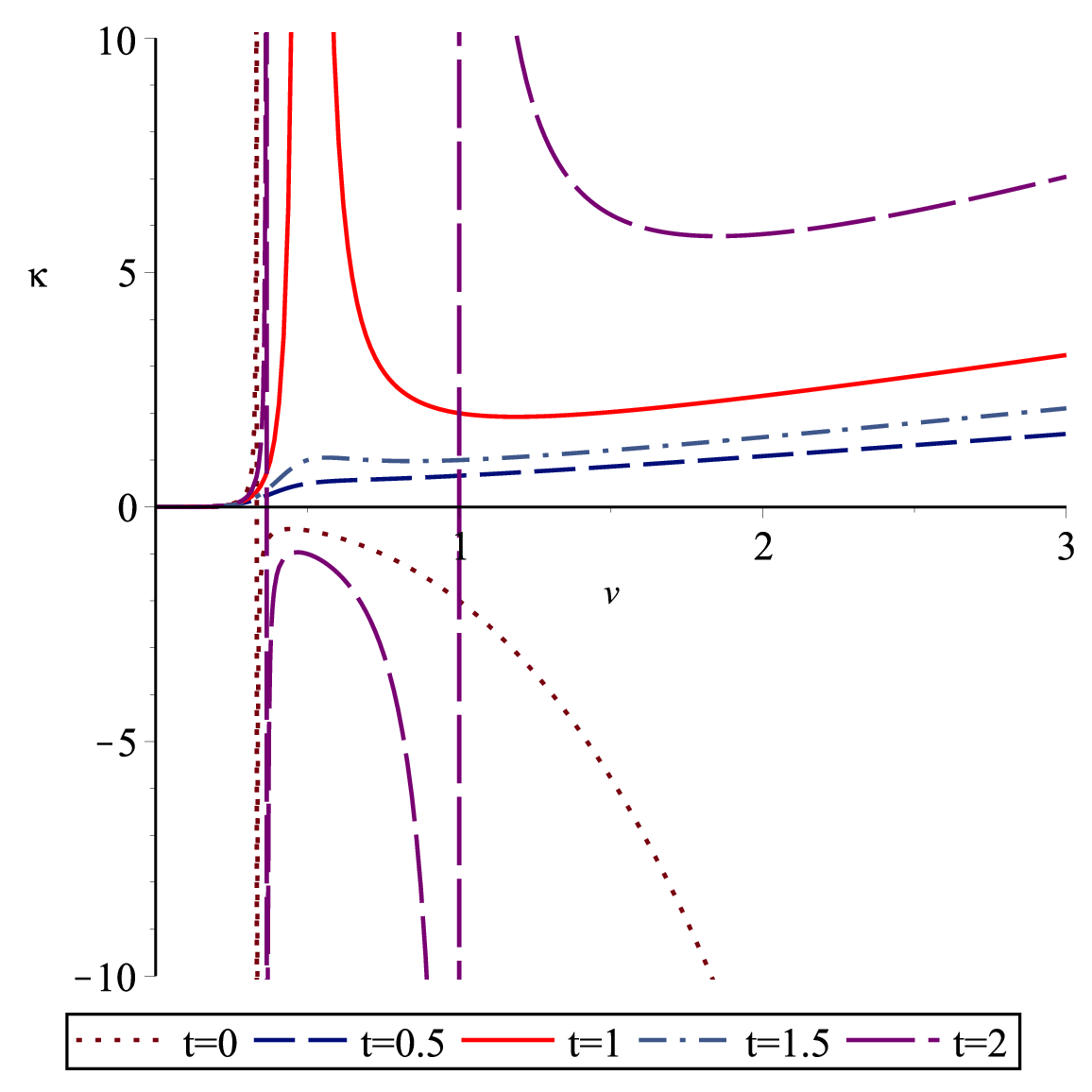}
\includegraphics[width=5cm]{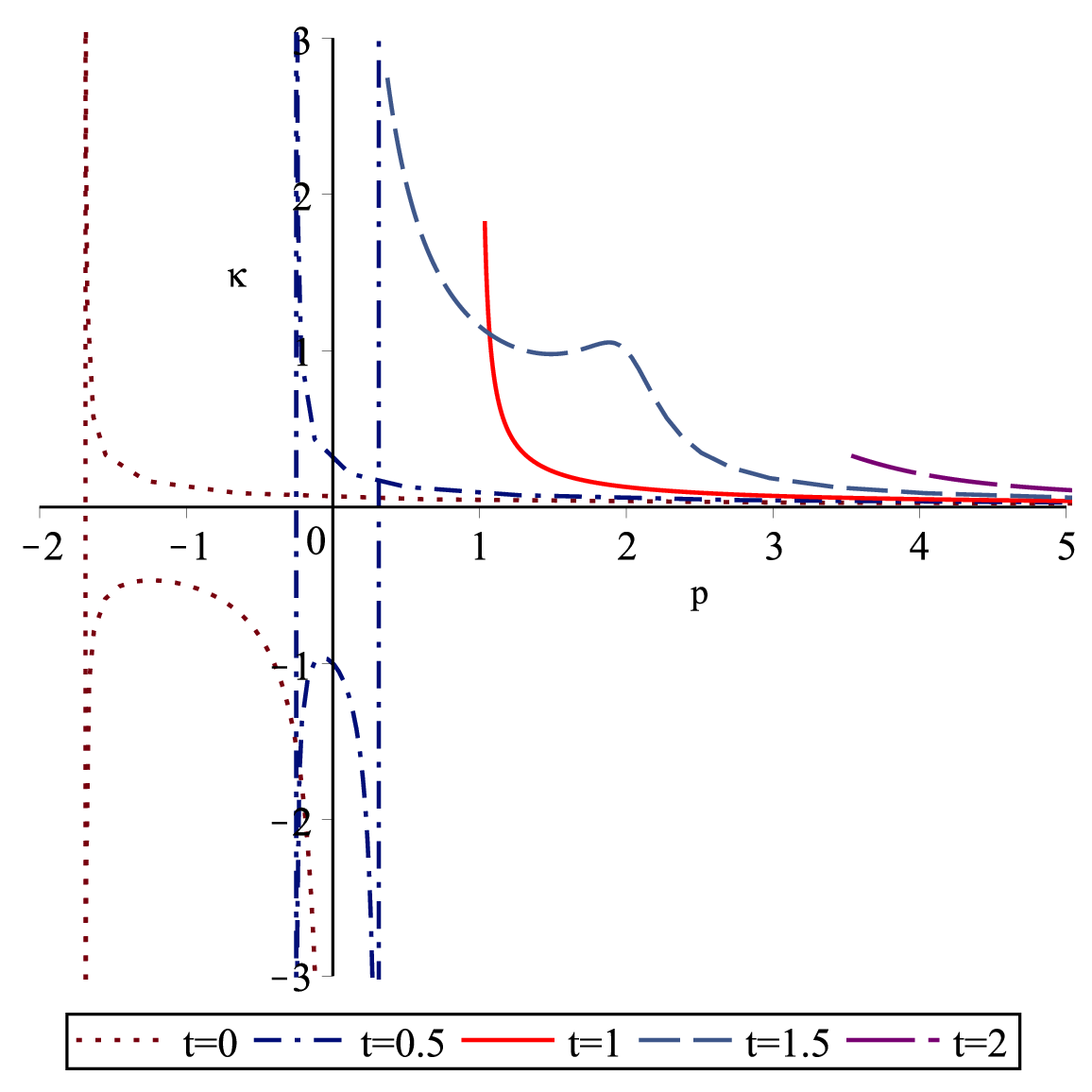}
\caption{\footnotesize{Diagrams for the coefficient of volume
expansion $\beta$ and compressibility coefficient $\kappa$ are
plotted vs the temperature at constant pressure and vs the
specific volume at constant pressure respectively. }}
\end{figure}

\begin{thebibliography}{99}
\bibitem{Will} C. M. Will, "The Confrontation between General Relativity and Experiment", Liv. Rev. Rel. 17, 4 (2014).
\bibitem{Berti} E. Berti, E. Barausse, V. Cardoso and et al, "Testing General Relativity with Present and Future Astrophysical
Observations", Class. Quantum Grav. 32, 243001 (2015).
\bibitem{Abbott} B. P. Abbott, "Observation of Gravitational Waves from a Binary Black Hole Merger", Phys. Rev. Lett. 118, 221101 (2017).
\bibitem{Hees} A. Hees,T. Do, A. M. Ghez and et al, "Testing General Relativity with stellar orbits around the supermassive black hole in our
 Galactic center", Phys. Rev. Lett. 118, 211101 (2017).
\bibitem{Heisenberg} L. Heisenberg, "A systematic approach to generalisations of General Relativity and their cosmological implications"
, Phys. Rept. 796, 1 (2019).
\bibitem{Clifton} T. Clifton, P. G. Ferreira, A. Padilla, C. Skordis, "Modified gravity and cosmology", Phys. Rept. 513, 1 (2012).
\bibitem{Weinberg} S. Weinberg,
"Gravitation and cosmology: Principals and Applications of the
General theory of relativity", John Wiley and Sons, New York
(1972).
\bibitem{Trimble} V Trimble, "Existence and nature of dark matter in the universe", Ann. Rev. Astron. Astrophys. 25, 425 (1987).
\bibitem{Moffat}J. W. Moffat, "Scalar-tensor-vector gravity theory", JCAP 0603, 004 (2006).
\bibitem{Moffat1} J. W. Moffat and S. Rahvar, "The MOG weak field approximation and observational test of galaxy rotation curves", Mon. Not. Roy. Astron. Soc. 436, 1439 (2013).
\bibitem{Ghaffar0} H.
Ghaffarnejad, "Scalar-vector-tensor gravity from preferred
reference frame effects", Gen, Relativ. Gravit. 40, 2229 (2008)
\bibitem{Ghaffar300} H. Ghaffarnejad, "Erratum to: Scalar-vector-tensor gravity
from preferred reference frame effects", Gen. relativ. Gravit, 41,
2941 (2009).
\bibitem{Brownstein} J.R. Brownstein and J.W. Moffat, "Galaxy cluster masses without non-baryonic dark matter", Mon. Not. Roy. Astron. Soc. 367
527, (2006).
\bibitem{Moffat2} J. W. Moffat and S. Rahvar, "The MOG weak field approximation
II:  Observational test of Chandra X-ray clusters", Mon. Not. Roy.
Astron. Soc. 441, 3724 (2014).
\bibitem{Moffat3} J.W. Moffat, "Scalar and vector field constraints, deflection of light and lensing in Modified Gravity (MOG)", arXiv:1410.2464 [gr-qc].
\bibitem{Moffat4} J.W. Moffat, "Structure growth and the CMB in Modified Gravity (MOG)", arXiv:1409.0853 [astroph.CO].
\bibitem{Ayon1} E. A. Beato and A. Garcia, "Regular Black Hole in General Relativity Coupled to Nonlinear Electrodynamics", Phys. Rev. Lett. 80, 5056 (1998).
\bibitem{Ayon2} E. A. Beato and  A. Garcia, "New regular black hole solution from nonlinear electrodynamics", Phys. Rev. Lett. B 464, 25 (1999).
\bibitem{Ayon3} E. A. Beato and  A. Garcia, "Non-singular charged black hole solution for non-linear source", Gen. Relativ. Gravit. 31, 629 (1999).
\bibitem{Ayon4} E. A. Beato and A. Garcia, "Four-parametric regular black hole solution", Gen. Relativ. Gravit. 37, 635 (2005).
\bibitem{CaiMi} X. C. Cai and Y. G. Miao, "Quasinormal modes of the generalized ABG STVG black hole in the scalar-tensor-vector gravity", arXiv:2008.04576v1 [gr-qc].
\bibitem{Ghaffar2010} H. Ghaffarnejad, "Quantum Cosmology with effects of a preferred reference frame", Class. Quantum Grav.
27,  015008 (2010).
\bibitem{Ghaffar2015} H. Ghaffarnejad, "Wave function of the Universe, preferred reference frame effects and metric signautre transition",
J. of Phys. 633, 012020 (2015).
 \bibitem{GhaffarDeh2019} H. Ghaffarnejad and R. Dehghani, "Galaxy rotation Curves and preferred reference frame effects",
 Eur. Phys. J. C79, 468 (2019).
 \bibitem{Ghaffar2019} H. Ghaffarnejad, "Canonical quantization of anisotropic Bianchi I cosmology from scalar vector tensor Brans Dicke gravity"
 , J. of Phys. 1391, 012028 (2019).
 \bibitem{Haw} S. W. Hawking, "Black Hole explosions?", Nature (London)248, 30, (1974)
 \bibitem{Haw1} S. W. Hawking, "Particle creation by black holes" Commun. Math, Phys. 43, 199 (1975).
\bibitem{Davies} P.C.W. Davies, "The thermodynamic theory of black holes", Proc. R. Soc. Lond. A 353, 499 (1977).
\bibitem{Hawking} S.W. Hawking, "Gravitational radiation from colliding black holes", Phys. Rev. Lett. 26, 1344 (1971).
\bibitem{Bardeenn} J. M. Bardeen, B. Carter and S. W. "Hawking, The Four laws of black hole mechanics", Commun. Math. Phys. 31, 161 (1973).
\bibitem{Bekenstein} J.D. Bekenstein, "Black holes and entropy", Phy. Rev. D 7, 2333 (1973)
\bibitem{Hawkingpage}S. W. Hawking and D. N. Page, "Thermodynamics of black holes in anti-de Sitter space", Commun. Math. Phys. 87, 577 (1983).
\bibitem{Chamblin1}A. Chamblin, R. Emparan, C.V. Johnson and R.C. Myers, "Charged AdS black holes and catastrophic holography"
, Phys. Rev. D 60, 064018 (1999).
\bibitem{Chamblin2}A. Chamblin, R. Emparan, C.V. Johnson and R.C. Myers, "Holography, thermodynamics, and fluctuations of charged AdS black holes"
, Phys. Rev. D 60, 104026 (1999).
\bibitem{Altamirano} N.
Altamirano, D. Kubiznak, R. B. Mann and Z. Sherkatghanad,
"Thermodynamics of rotating black holes and black rings: phase
transitions and thermodynamic volume", Galaxies 2, 89 (2014).
\bibitem{Chen} S. Chen, X. Liu, C. Liu, J. Jing, "P-V criticality of AdS black hole in f(R) gravity", Chin. Phys. Lett. 30,  060401 (2013).
\bibitem{Zou1} D.C. Zou, S.J. Zhang, B. Wang, "Critical behavior of Born-Infeld AdS black holes in the extended phase space thermodynamics"
, Phys. Rev. D 89, 044002 (2014) .
\bibitem{Cai2} R.G. Cai, L.M. Cao, L. Li, R.Q. Yang, "P-V criticality in the extended phase space of Gauss-Bonnet black holes in AdS space"
, JHEP. 1309, 005 (2013); arXiv:1306.6233 [gr-qc].
\bibitem{Wei} S.W. Wei, Y.X. Liu, "Triple points and phase diagrams in the extended phase space of charged Gauss-Bonnet black holes in AdS space"
    Phys. Rev. D 90, 044057 (2014), arXiv:1402.2837.
\bibitem{Zou2} D.C. Zou, Y. Liu, B. Wang, "Critical behavior of charged Gauss-Bonnet AdS black holes in the grand canonical ensemble"
, Phys. Rev. D. 90, 044063 (2014); arXiv:1404.5194.
\bibitem{Dehghani} M.H. Dehghani, S. Kamrani, A. Sheykhi, "P-V criticality of charged dilatonic black holes"
, Phys. Rev. D 90, 104020 (2014); arXiv:1505.02386 [hep-th].
\bibitem{Stetsko1} M. M. Stetsko, "Static spherically symmetric Einstein-Yang-Mills-dilaton black hole
and its thermodynamics", Phys. Rev. D 101, 124017  (2020);
arXiv:2005.13447v1 [hep-th].
\bibitem{Stetsko2} M. M. Stetsko, "Static spherically symmetric Einstein-Maxwell-Yang-Mills-dilaton black
hole and its thermodynamics", arXiv:2007.00277v1 [hep-th].
\bibitem{Dolan1} B.P. Dolan, A. Kostouki, D. Kubiznak, R.B. Mann, "Isolated critical point from Lovelock gravity",
 Class. Quantum Grav. 31, 24, 242001 (2014); arXiv:1407.4783 [hep-th].
\bibitem{Frassino} A.M. Frassino, D. Kubiznak, R.B. Mann, F. Simovic, "Multiple Reentrant Phase Transitions and Triple Points in Lovelock Thermodynamics"
, JHEP. 1409, 080 (2014); arXiv:1406.7015 [hep-th].
\bibitem{Xu} H. Xu, W. Xu, L. Zhao, "Extended phase space thermodynamics for third
order Lovelock black holes in diverse dimensions", Eur. Phys. J. C
74(9) (2014) 3074, arXiv:1405.4143 [gr-qc].
\bibitem{Gunasekaran} S. Gunasekaran, D. Kubiznak, R.B. Mann,
"Extended phase space thermodynamics for charged and rotating
black holes and Born-Infeld vacuum polarization", JHEP 1211, 110
(2012)' arXiv:1208.6251 [hep-th].
\bibitem{Belhaj} A. Belhaj, M. Chabab, H. El Moumni, M.B. Sedra, "On thermodynamics of AdS black holes in arbitrary dimensions"
, Chin. Phys. Lett. 29, 100401
 (2012) .
\bibitem{Dolan2} B.P. Dolan, "Thermodynamic stability of asymptotically anti-de Sitter rotating black holes in higher dimensions", Class. Quantum Grav. 31 165011 (2014);
arXiv:1403.1507.
\bibitem{Ghaffar2} H. Ghaffarnejad, "Stability of the evaporating Schwarzschild-de Sitter black hole final state", Phys. Rev. D74, 104012 (2006).
\bibitem{Ghaffar3} H. Ghaffarnejad, "Quantum field back reaction corrections and remnant stable evaporating Schwarzschild de Sitter dynamical black
hole", Phys. Rev. D75, 084009 (2007).
\bibitem{Ghaffar4} H. Ghaffarnejad, H. Neyad and M. A. Mojahedi,"Evaporating quantum Lukewarm black holes final state from back
reaction corrections of quantum scalar fields
", Astrophys. and Space Sci, 346, 497, (2013).
 \bibitem{Ghaffar5} H. Ghaffarnejad, "Classical and Quantum Reissner-Nordstrom Black Hole Thermodynamics and first order Phase
 Transition", Astrophys. and Space Sci.361,7,6  (2016),arXiv:1308.1323 [physics.gen-ph].
\bibitem{Ghaffar6} H. Ghaffarnejad and M. Farsam, "Reissner-Nordstrom black holes statistical ensembles and first order thermodynamic phase
transition", Advances in High Energy Physics, 2019, 2539217
(2019), arXiv:1603.08408 [physics.gen-ph].
\bibitem{Ghaffar7} H. Ghaffarnejad, E. Yaraie and M. Farsam, "Quintessence Reissner Nordstrom Anti de Sitter Black Holes and Joule Thomson
effect", Int. J. Theor. Phys. 57, 1671, (2018); arXiv:1802.08749
[gr-qc]
\bibitem{Ghaffar8} H. Ghaffarnejad, M. Farsam and E. Yaraie, "Holographic entanglement entropy for small subregions and thermalization of Born-Infeld AdS black
holes", Nucl. Phys. B938, 523 (2019);     arXiv:1803.05725
[hep-th]
\bibitem{Ghaffar9} H. Ghaffarnejad, M. Farsam and E. Yaraie, "Effects of quintessence dark energy on the action growth and butterfly
velocity",     Advances in High Energy Physics  2020, 9529356,
(2020),     arXiv:1806.05735 [hep-th],
\bibitem{Ghaffar10} H. Ghaffarnejad, E. Yaraie and M. Farsam, "Holographic thermalization in AdS-Gauss-Bonnet gravity for small entangled
regions", Gen. Relativ. Gravit.51, 10 (2019);     arXiv:1806.05976
[hep-th] \bibitem{Ghaffar11} H. Ghaffarnejad and E. Yaraie,
"Effects of a cloud of strings on the extended phase space of
Einstein-Gauss-Bonnet AdS black holes", Phys. Let. B785, 10,105
(2018);     arXiv:1806.06687 [gr-qc]
\bibitem{Ghaffar12} H. Ghaffarnejad and M. Farsam, "The Last Lost Charge And Phase Transition In Schwarzschild AdS Minimally Coupled to a Cloud of
Strings", Eur. Phys. J. Plus, 134, 110 (2019);
arXiv:1806.06688 [hep-th].
\bibitem{Ghaffar13} E. Yaraie, H. Ghaffarnejad and M. Farsam, "Complexity growth and shock wave geometry in AdS-Maxwell-power-Yang-Mills
theory", Eur. Phys. J. C78, 967 (2018);     arXiv:1806.07242
[gr-qc]
\bibitem{Ghaffar14} H. Ghaffarnejad, E. Yaraie and M. Farsam, "Thermodynamic phase transition for Quintessence Dyonic Anti de Sitter Black
Holes", Eur. Phys. J. Plus, 135, 179 (2020);     arXiv:1808.09789
[physics.gen-ph]
\bibitem{Ghaffar15} H. Ghaffarnejad, E. Yaraie, M. Farsam and K.
Bamba, "Hairy black holes and holographic heat engine", Nucl.
Phys. B, 952, 114941, (2020);     arXiv:1809.10142 [hep-th]
\bibitem{Ghaffar16} M. Farsam, E. Yaraie, H. Ghaffarnejad and E.
Ghasami, "Cooling-heating phase transition for 4D AdS Bardeen
Gauss-Bonnet Black Hole",     arXiv:2010.05697 [hep-th]
\bibitem{Ghaffar17} H. Ghaffarnejad, E. Yaraie and M. Farsam, "4D Gauss-Bonnet Black Holes in AdS Space Surrounded by Strings Fluid Mimics Van der
Waals Fluid Behavior"
    arXiv:2010.07108 [hep-th]
\bibitem{Moffat5} J. W. Moffat, "Black holes in modified gravity (MOG)", Eur. Phys. J. C 75,  175 (2015).
\bibitem{Haydarov} K. Haydarov, J.Rayimbaev, A. Abdujabbarov, S. Palvanov and D.
Begmatova, "Magnetized particle motion around magnetized
Schwarzschild-MOG black hole", Eur. Phys. J. C80, 399, (2020),
\bibitem{Mureika} J. R. Mureika,
J. W. Moffat and M. Faizal, "Black hole thermodynamics in Modified
Gravity(MOG)", Phys. Lett. B757, 528 (2016).
\bibitem{Kastor} D. Kastor, S. Ray and J. Traschen, "Enthalpy and the Mechanics of AdS Black Holes",
Class. Quantum Grav.26, 195011 (2009) , [arXiv:0904.2765
[hep-th]].
 \bibitem{Dolan} B. P. Dolan, "Pressure and volume in the first law of black hole
 thermodynamics", Class. Quantum Grav. 28, 235017 (2011) [1106.6260v3 [gr-qc]].
\bibitem{Malda} J. M.
Maldacena, "The large N limit of superconformal field theories and
supergravity," Int. J. Theor. Phys.38, 1113 (1999);
[arXiv:hep-th/9711200]
\bibitem{Gubser}  S. S. Gubser, I. R. Klebanov and A. M. Polyakov, "Gauge theory correlators from  non-critical string theory,"
Phys. Lett. B428, 105 (1998) [arXiv:hep-th/9802109].
\bibitem{Witten} E. Witten, "Anti-de Sitter space and holography," Adv. Theor. Math. Phys.2, 253(1998) [arXiv:hep-th/9802150].
\bibitem{spergel}  D. N. Spergelet et al. [WMAP Collaboration], "Wilkinson Microwave Anisotropy Probe(WMAP)
three year results: Implications for cosmology," Astrophys. J.
Suppl.170,377 (2007) [arXiv:astro-ph/0603449]
\bibitem{carter} B.
Carter, "Hamilton-Jacobi and Schrodinger separable solutions of
Einstein equations," Commun. Math. Phys.10, 280 (1968)
\bibitem{Don} Don N. Page, "Hawking radiation and black hole thermodynamics", New J. Phys.7.
203, (2005).
\bibitem{Smarr} L. Smarr, "Mass  Formula  for  Kerr  BlackHoles", Phys. Rev. Lett3(1972) 71,  Erratum  ibid.,30(1973) 521
\bibitem{Shi} Sh. Q. Hu, X. M. Kuang and Y. Ch. Ong, "A Note on Smarr Relation and Coupling
Constants", Gen Relativ Gravit. 51, 55 (2019) [ arXiv:1810.06073
[gr-qc]].
\end{thebibliography}
\end{document}